\documentclass[12pt]{article}

\usepackage{amssymb}
\usepackage{amsmath}
\usepackage{amscd}
\usepackage{mathrsfs}
\usepackage{amssymb}
\usepackage{feynmf}
\usepackage{graphicx}

\def\dj{\hbox{d\kern-0.347em \vrule width 0.3em height 1.252ex depth
-1.21ex \kern 0.051em}}

\numberwithin{equation}{section}

\begin{document}

\setlength{\oddsidemargin}{0cm}
\setlength{\baselineskip}{7mm}


\thispagestyle{empty}
\setcounter{page}{0}

\begin{flushright}

\end{flushright}

\vspace*{1cm}

\begin{center}
{\bf \Large The Double Copy Structure of Soft Gravitons}

\vspace*{0.3cm}

\vspace*{1cm}

Agust\'{\i}n Sabio Vera$^{a,}$\footnote{\tt 
a.sabio.vera@gmail.com}
and Miguel A. V\'azquez-Mozo$^{b,}$\footnote{\tt 
Miguel.Vazquez-Mozo@cern.ch}

\end{center}

\vspace*{0.0cm}

\begin{center}

$^{a}${\sl Instituto de F\'{\i}sica Te\'orica UAM/CSIC \&
Universidad Aut\'onoma de Madrid\\
C/ Nicol\'as Cabrera 15, E-28049 Madrid, Spain
}

$^{b}${\sl Departamento de F\'{\i}sica Fundamental \& IUFFyM \\
 Universidad de Salamanca \\ 
 Plaza de la Merced s/n,
 E-37008 Salamanca, Spain
  }
\end{center}

\vspace*{1.5cm}

\centerline{\bf \large Abstract}

\noindent
The subleading corrections to factorization theorems for soft bremsstrahlung in nonabelian 
gauge theories and gravity are investigated in the case of a five point amplitude with four scalars. Building on recent results, we write the action of 
the angular momentum operators on scattering amplitudes as derivatives with respect to 
the Mandelstam invariants to uncover a double copy structure in the contribution of the soft 
graviton to the amplitude, both in the leading term and the first correction. Using our approach, we study 
Gribov's theorem as extended to nonabelian gauge theories and gravity by Lipatov,  
and find that subleading corrections can be obtained from those to Low's theorem by dropping the terms with derivatives with respect to the center-of-mass energy, which are suppressed at high energies. In this case, the emitted gravitons are not necessarily soft.

\newpage  

\setcounter{footnote}{0}

\section{Introduction}

The scattering amplitudes of quantum field theories with massless intermediate gauge bosons have an interesting infrared behavior, in particular in the soft limit where massless bosons are emitted with very small momenta. In this context, it was proved by Low~\cite{low} that in QED the leading
behavior of an inelastic amplitude with an emitted soft photon is dominated by those contributions in which the photon is {\em bremsgestrahlt} by the external states (Low formulated the theorem for scalar charged particles, Burnett and Kroll generalized it to the case of charged fermions~\cite{Burnett:1967km}). Subleading corrections to this result, sensitive to internal emissions, were also computed in~\cite{low} and found
to have a particularly simple form. 
In the case of gravity, Weinberg showed~\cite{weinberg} that a similar result holds for scattering amplitudes in which a soft graviton is emitted. 

More recently, there has been a renewed interest in these results stemming from the realization that Weinberg's soft-graviton theorem can 
be regarded as the Ward identity associated with the symmetries of the gravitational theory at null 
infinity~\cite{strominger_et_al}. 
This has led to the formulation of a 
new soft-graviton theorem including next-to-leading and next-to-next-to-leading order corrections which have a universal expression in terms of the
angular momentum of the hard particles~\cite{cachazo_strominger}. Using obvious notation,
\begin{eqnarray}
\mathcal{M}_{n+1}(k;p_{1},\ldots,p_{n})&=&\kappa\left[\sum_{i=1}^{n}{\varepsilon^{\mu\nu}p_{i\mu}p_{i\nu}\over p_{i}\cdot k}+
\sum_{i=1}^{n}{\varepsilon^{\mu\nu}p_{i\mu}(k^{\alpha}J^{(i)}_{\nu\alpha})\over p_{i}\cdot k} \right.\nonumber \\[0.2cm]
&+&\left.\sum_{i=1}^{n}{\varepsilon^{\mu\nu}(k^{\alpha}J^{(i)}_{\mu\alpha})(k^{\beta}J^{(i)}_{\nu\beta})\over p_{i}\cdot k}\right]
\mathcal{M}_{n}(p_{1},\ldots,p_{n}),
\end{eqnarray}
where 
\begin{eqnarray}
J_{\mu\nu}^{(i)}=p_{i \mu}{\partial\over\partial p_{i}^{\nu}}-p_{i \nu}{\partial\over\partial p_{i}^{\mu}}
\end{eqnarray}
is the angular momentum operator of the $i$-th particle.
It has been argued that this result is not renormalized~\cite{cachazo_yuan,bianchi_et_al}. Generalizations of
this new soft-graviton theorem to arbitrary dimensions were studied in \cite{higher-dimensions}.

In the case of Yang-Mills theories, the subleading corrections to Low's result can also be encoded in terms of the angular momentum operator acting on the $n$-point amplitude~\cite{casali,schwab_volovich}:
\begin{eqnarray}
\mathcal{A}_{n+1}(k;p_{1},\ldots,p_{n})&=& g\left(\sum_{i=1}^{n}{\epsilon\cdot p_{i}\over p_{i}\cdot k}+\sum_{i=1}^{n}
{\epsilon^{\mu}k^{\nu}J^{(i)}_{\mu\nu}\over p_{i}\cdot k}\right)\mathcal{A}_{n}(p_{1},\ldots,p_{n}).
\label{eq:low+subleading_general}
\end{eqnarray}
In QED, the subleading corrections also admit an interpretation in terms of the asymptotic symmetries of the theory at null infinity~\cite{lysov_et_al}.

Low's theorem was originally derived in the limit in which the momentum of the photon is taken to be very small, $k\rightarrow 0$. It was later realized by Gribov~\cite{gribov} that the expression found by Low has a broader range of validity if the scattering takes place at a large center-of-mass energy of the colliding particles, $\sqrt{s}$. In this case the factorization can also hold for hard emissions as long as their  transverse momentum with respect to the radiating particle is small compared to the momentum transfers typical of the scattering process. Thus, for two colliding hadrons of typical mass $\mu$ and momenta 
$p$ and $q$, the amplitude is dominated by external bremsstrahlung in the kinematic region defined by
\begin{eqnarray}
{2\,p\cdot k}, \, \,  {2\, q\cdot k}\ll s \hspace*{1cm} \mathbf{k}_{\perp}^{2}\approx{(2 \, p\cdot k)(2\,q\cdot k)\over s}
\ll\mu^{2},
\label{eq:gribov_region}
\end{eqnarray}
with $\mathbf{k}_{\perp}$ the transverse momentum of the photon. The main difference with respect to the regime of validity 
of Low's theorem (which applies in the region $2 \, p \cdot k, 2 \, q \cdot k \ll \mu^2$) is that now we assume a large center-of-mass energy $\sqrt{s} \gg \mu$, without requiring the photon momentum to be soft. 
The theorem was generalized by Lipatov in~\cite{lipatov1} to the case of a Yang-Mills field or a graviton coupled to scalars. 

In this note we study the corrections to soft gluon and graviton theorems for amplitudes containing scalar fields, and investigate the double copy structure
of the latter one. In Section~\ref{sec:soft_QCD},
we generalize the analysis of~\cite{BDDVN} to the nonabelian case for the scattering amplitude of two different scalars with the emission of a gluon,
showing that the first correction to Low's leading result is completely fixed by gauge invariance, as it happens in QED. 
Writing the action of the angular momentum operators on the four scalars amplitude using derivatives with respect to the Mandelstam $s$ and $t$ invariants, we find
that the amplitude has a particularly simple form in terms of a set of gauge invariant coefficients. Due to the Jacobi identity satisfied by 
the color factors, these coefficients admit shift transformations that preserve the value of the amplitude and do play an important role when connecting the gauge theory amplitude to the corresponding gravitational one.  In Section~\ref{sec:grav_amplitu_dc} we analyze the gravitational scattering amplitude of two scalars with the emission of a graviton in the soft limit. 
As in the gauge theory case, we express the first correction in terms of derivatives with respect to the Mandelstam invariants and find that the associated 
coefficients have a double copy structure. This is interpreted in the sense that the 
contribution of the soft graviton to the five-point gravitational amplitude can be factored as the square of the contribution 
of the soft gluon to its gauge theory counterpart, after removing the color factors. Section~\ref{sec:gribov} is devoted to the study of the Gribov limit, which allows for not necessarily soft  bremsstrahlung, both in gauge theories and 
gravity. We find that the first correction to the amplitude in this kinematic region computed in \cite{lipatov1}
can be obtained from the corresponding correction to the Low/Weinberg
theorems by dropping derivatives with respect to the $s$ Mandelstam invariant. Finally, in Section~\ref{sec:conclusions} we summarize our conclusions.

\section{Soft gluons in scalar QCD}
\label{sec:soft_QCD}

\begin{figure}
\centerline{\includegraphics[scale=0.8]{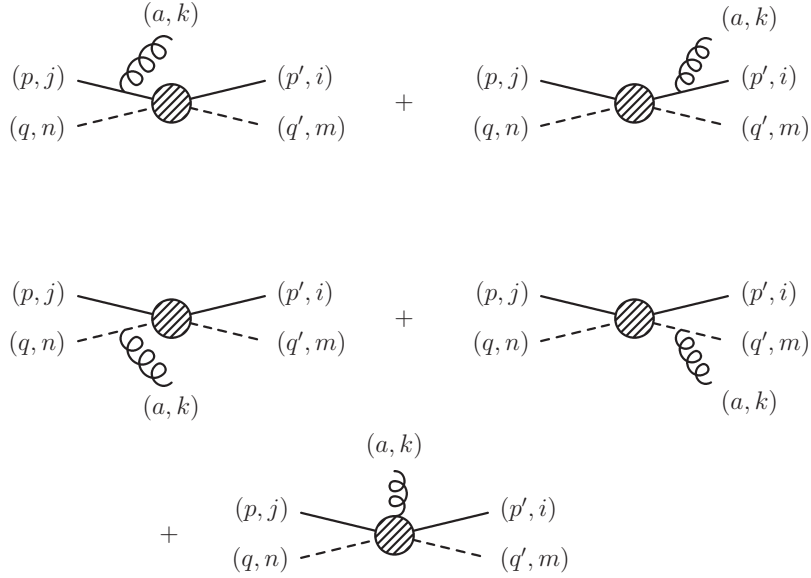}}
\caption[]{Generic topologies contributing to the scattering of two distinct scalars with gluon emission. The momenta $p$ and $q$ are taken
incoming, while $k$, $p'$, and $q'$ are outgoing.}
\label{fig:topologies}
\end{figure}

It has been shown in~\cite{BDDVN} that in scalar QED the first subleading correction to Low's theorem is completely fixed by gauge invariance. In this 
section we extend this result to the nonabelian case by considering scalar QCD (sQCD) with two flavors and the scattering amplitude of two distinct scalars in an 
arbitrary representation with radiation of a gluon. 
The five generic topologies contributing to this process are shown in Fig.~\ref{fig:topologies}: 
four of them correspond to the bremsstrahlung of a gluon
by the external scalars, while in the fifth one the gluon is emitted from an internal propagator. Based on Lorentz and color covariance, the amplitude can be written as
\begin{eqnarray}
\mathcal{A}_{5}&=&2g\left[c_{1}{p'\cdot\epsilon\over s_{1'}}\mathcal{A}_{4}(p,q,p'+k,q')-c_{2}{p\cdot\epsilon\over s_{1}}
\mathcal{A}_{4}(p-k,q,p',q')\right. \nonumber \\[0.2cm]
&+&\left.c_{4}{q'\cdot \epsilon\over s_{2'}}\mathcal{A}_{4}(p,q,p',q'+k)-c_{5}{q\cdot \epsilon\over s_{2}}
\mathcal{A}_{4}(p,q-k,p',q')\right] \label{eq:general_amplitude_proof}  \\[0.2cm]
&+& {g\over 2}\Big[c_{3} \, \epsilon_{\mu}\mathcal{B}_{1}^{\mu}(k;p,q,p',q')
+c_{6} \, \epsilon_{\mu}\mathcal{B}_{2}^{\mu}(k;p,q,p',q')
+c_{7} \, \epsilon_{\mu}\mathcal{B}_{3}^{\mu}(k;p,q,p',q')\Big].
\nonumber
\end{eqnarray}
Here $\mathcal{A}_{4}(p,q,p',q')$ is the color-stripped four-point scalar amplitude, while the three functions
$\mathcal{B}^{\mu}_{1}(k;p,q,p',q')$, $\mathcal{B}^{\mu}_{2}(k;p,q,p',q')$, and $\mathcal{B}^{\mu}_{3}(k;p,q,p',q')$ 
parametrize the diagrams with internal emissions. The
color factors are given by~\cite{SVSCVM}
\begin{eqnarray}
c_{1}&=& T^{a}_{ik}T^{b}_{kj}\widetilde{T}^{b}_{mn}, \hspace*{4cm}  c_{5}= T^{b}_{ij}\widetilde{T}^{b}_{m\ell}\widetilde{T}^{a}_{\ell n}, \nonumber \\[0.2cm]
c_{2}&=& T^{b}_{ik}T^{a}_{kj}\widetilde{T}^{b}_{mn},  \hspace*{4cm}  
c_{6}= T^{b}_{ij}\widetilde{T}^{a}_{m\ell}\widetilde{T}^{b}_{\ell n}+T^{b}_{ij}\widetilde{T}^{b}_{m\ell}\widetilde{T}^{a}_{\ell n}, \nonumber \\[0.2cm]
c_{3}&=& T^{a}_{ik}T^{b}_{kj}\widetilde{T}^{b}_{mn}+T^{b}_{ik}T^{a}_{kj}\widetilde{T}^{b}_{mn} , \hspace*{1.7cm} 
c_{7}= if^{abc}T^{b}_{ij}\widetilde{T}^{c}_{mn}, \label{eq:color_factors_scalar} \\[0.2cm]
c_{4}&=& T^{b}_{ij}\widetilde{T}^{a}_{mk}\widetilde{T}^{b}_{kn},   \nonumber
\end{eqnarray}
where $T_{ij}$ and $\widetilde{T}_{mn}$ are the gauge group generators associated with the two scalar flavors. 
We have also introduced the following kinematic invariants
\begin{eqnarray}
s_{1}=2 \, k\cdot p, \hspace*{1cm} s_{2}=2 \, k\cdot q, \hspace*{1cm} s_{1'}=2 \, k\cdot p', \hspace*{1cm} s_{2'}=2 \, k\cdot q'.
\end{eqnarray}
The next step is to enforce gauge invariance. 
The gauge Ward identity reads,
\begin{eqnarray}
c_{1} \, \mathcal{A}_{4}(p,q,p'+k,q')-c_{2} \, \mathcal{A}_{4}(p-k,q,p',q')
+c_{4} \, \mathcal{A}_{4}(p,q,p',q'+k)-c_{5} \, \mathcal{A}_{4}(p,q-k,p',q') \nonumber \\[0.2cm]
+\,\,\,{k_{\mu} \over 2} \Bigg[c_{3} \,\mathcal{B}_{1}^{\mu}(k;p,q,p',q')+c_{6} \, \mathcal{B}_{2}^{\mu}(k;p,q,p',q')
+c_{7} \, \mathcal{B}_{3}^{\mu}(k;p,q,p',q') \Bigg]=0.\hspace*{0.5cm}
\end{eqnarray}
In the soft limit we expand this equation in powers of the gluon momentum. At leading order in this expansion, the Ward identity is
automatically satisfied due to the Jacobi identity $c_{1}-c_{2}+c_{4}-c_{5}=0$. In the linear approximation, on the other hand,
we are led to the equation
\begin{eqnarray}
k_{\mu} \Bigg[ 2 \left(c_{1}{\partial\over\partial p'^{\mu}}+c_{2}{\partial\over\partial p^{\mu}}
+c_{4}{\partial\over\partial q'^{\mu}}+c_{5}{\partial\over\partial q^{\mu}}\right)\mathcal{A}_{4}(p,q,p',q') \hspace*{5cm}
\nonumber \\[0.2cm]
+c_{3} \, \mathcal{B}_{1}^{\mu}(0;p,q,p',q')
+c_{6} \, \mathcal{B}_{2}^{\mu}(0;p,q,p',q')
+c_{7} \, \mathcal{B}_{3}^{\mu}(0;p,q,p',q') \Bigg]=0.
\label{eq:gauge_ward_nlo}
\end{eqnarray}
To solve for the functions $\mathcal{B}_{i}^{\mu}(0;p,q,p',q')$, we have to keep in mind that the color factors are not independent. In fact, there are
four independent Jacobi identities relating them, 
\begin{eqnarray}
c_{1}+c_{2}-c_{3}=0, &\hspace*{1cm}&
c_{4}+c_{5}-c_{6}=0, \\[0.2cm]
c_{1}-c_{2}+c_{7}=0, &\hspace*{1cm}&
c_{4}-c_{5}-c_{7}=0,
\nonumber
\end{eqnarray}
which can be used to eliminate $c_{1}$, $c_{2}$, $c_{4}$ and $c_{5}$ in favor of $c_{3}$, $c_{6}$ and 
$c_{7}$. Using these relations in Eq. \eqref{eq:gauge_ward_nlo} we arrive at
\begin{eqnarray}
k^{\mu} \Bigg\{c_{3} \, \left[{\partial\mathcal{A}_{4}\over\partial p'^{\mu}}+{\partial\mathcal{A}_{4}\over\partial p^{\mu}}
+\mathcal{B}_{1 \mu}(0;p,q,p',q')
\right]+c_{6} \, \left[{\partial\mathcal{A}_{4}\over\partial q'^{\mu}}+{\partial\mathcal{A}_{4}\over\partial q^{\mu}}
+\mathcal{B}_{2 \mu}(0;p,q,p',q')
\right] \nonumber \\[0.2cm]
+\,\,\,c_{7} \, \left[-{\partial\mathcal{A}_{4}\over\partial p'^{\mu}}+{\partial\mathcal{A}_{4}\over\partial p^{\mu}}
+{\partial\mathcal{A}_{4}\over\partial q'^{\mu}}-{\partial\mathcal{A}_{4}\over\partial q^{\mu}}
+\mathcal{B}_{3 \mu}(0;p,q,p',q')
\right] \Bigg\}=0.
\end{eqnarray}
Since $c_{3}$, $c_{6}$ and $c_{7}$ are independent, we arrive at the equations to obtain the three undetermined functions that are solved by
\begin{eqnarray}
\mathcal{B}_{1}^{\mu}(0;p,q,p',q')&=&-{\partial\mathcal{A}_{4}\over\partial p'^{\mu}}-{\partial\mathcal{A}_{4}\over\partial p^{\mu}}
, \nonumber \\[0.2cm]
\mathcal{B}_{2}^{\mu}(0;p,q,p',q')&=&-{\partial\mathcal{A}_{4}\over\partial q'^{\mu}}-{\partial\mathcal{A}_{4}\over\partial q^{\mu}}
, \label{eq:B's_solns} \\[0.2cm]
\mathcal{B}_{3}^{\mu}(0;p,q,p',q')&=&{\partial\mathcal{A}_{4}\over\partial p'^{\mu}}-{\partial\mathcal{A}_{4}\over\partial p^{\mu}}
-{\partial\mathcal{A}_{4}\over\partial q'^{\mu}}+{\partial\mathcal{A}_{4}\over\partial q^{\mu}}.
\nonumber
\end{eqnarray}
To these solutions for $\mathcal{B}^{\mu}_{i}(0;p,q,p',q')$ we could add a function $\Delta_{i}^{\mu}$ satisfying $k_{\mu}\Delta^{\mu}_{i}=0$. However, its
tensor structure implies that such 
a function must be at least linear in $k$ and therefore can be ignored at this order. 

Having calculated the leading behavior of the functions associated with the internal emission diagrams, 
we expand the five-point amplitude~\eqref{eq:general_amplitude_proof} to $ {\cal O}(k^{0})$ and substitute the expressions found in~\eqref{eq:B's_solns}. Using the Jacobi identities
to eliminate the color factors $c_{3}$, $c_{6}$, and $c_{7}$ we have
\begin{eqnarray}
\mathcal{A}_{5}&=&2g \Bigg[c_{1}{p'\cdot\epsilon\over s_{1'}}-c_{2}{p\cdot\epsilon\over s_{1}}
+c_{4}{q'\cdot \epsilon\over s_{2'}}-c_{5}{q\cdot \epsilon\over s_{2}}
 \nonumber \\[0.2cm] 
&+& k^{\mu}\left(c_{1}{p'\cdot\epsilon\over s_{1'}}{\partial \over\partial p'^{\mu}}+
c_{2}{p\cdot\epsilon\over s_{1}}{\partial \over\partial p^{\mu}}
+c_{4}{q'\cdot \epsilon\over s_{2'}}{\partial \over\partial q'^{\mu}}
+c_{5}{q\cdot \epsilon\over s_{2}}{\partial \over\partial q^{\mu}}\right) \\[0.2cm]
&-& \frac{1}{2} \epsilon^{\mu}\left(c_{1}{\partial\over\partial p'^{\mu}}+c_{2}{\partial\over\partial p^{\mu}}
+c_{4}{\partial\over \partial q'^{\mu}}+c_{5}{\partial\over\partial q^{\mu}}\right)
\Bigg]\mathcal{A}_{4}(p,q,p',q')
+\mathcal{O}(k).
\nonumber
\end{eqnarray}
After a few manipulations, this can be recast in terms of the angular momentum operators as
\begin{eqnarray}
\mathcal{A}_{5}&=&2g \Bigg(c_{1}{p'\cdot\epsilon\over s_{1'}}-c_{2}{p\cdot\epsilon\over s_{1}}
+c_{4}{q'\cdot \epsilon\over s_{2'}}-c_{5}{q\cdot \epsilon\over s_{2}}\nonumber
\label{eq:amplitude_5p_general} \\[0.2cm]
&+&c_{1}{\epsilon^{\mu}k^{\nu}J^{(1')}_{\mu\nu}\over s_{1'}}+c_{2}{\epsilon^{\mu}k^{\nu}J^{(1)}_{\mu\nu}\over s_{1}}
+c_{4}{\epsilon^{\mu}k^{\nu}J^{(2')}_{\mu\nu}\over s_{2'}}+c_{5}{\epsilon^{\mu}k^{\nu}J^{(2)}_{\mu\nu}\over s_{2}} \Bigg)
\mathcal{A}_{4}(p,q,p',q'). \hspace*{0.5cm} 
\end{eqnarray}

We have shown that the first correction to the amplitude in the soft limit is completely fixed by the requirement of gauge invariance.  
Compared to the scalar QED case analyzed in~\cite{BDDVN}, we have a larger number of unknown functions associated with the different color structures
in the internal emission diagrams. However, this very fact implies that there is an equally larger number of 
independent constraints to determine these functions. 
At linear order in the gluon momentum, we have again three equations for the first derivatives of $\mathcal{B}_{i}(k;p,q,p',q')$ at $k=0$, but as
in the Abelian case these relations leave the curls
\begin{eqnarray}
\left.\left({\partial\mathcal{B}_{i}^{\mu}\over\partial k_{\alpha}}-{\partial\mathcal{B}_{i}^{\alpha}\over\partial k_{\mu}}\right)\right|_{k=0},
\hspace*{1cm} i=1,2,3,
\end{eqnarray}
undetermined. 

The corrections to Low's theorem in Eq.~\eqref{eq:amplitude_5p_general} can be rewritten using the Mandelstam invariants $s$ and $t$, that we define in the following symmetric form
\begin{eqnarray}
s&=&{1\over 2}(p+q)^{2}+{1\over 2}(p'+q')^{2}, \nonumber \\[0.2cm]
t&=&{1\over 2}(p-p')^{2}+{1\over 2}(q-q')^{2}.
\label{eq:mandelstam_inv}
\end{eqnarray}
The combinations containing the angular momentum operators can then be expressed in terms of derivatives with respect to $s$ and $t$ as
\begin{eqnarray}
\epsilon^{\mu}k^{\nu}J^{(1')}_{\mu\nu}&=&A_{1'}{\partial\over\partial s}+B_{1'}{\partial\over\partial t},
\nonumber \\[0.2cm]
\epsilon^{\mu}k^{\nu}J^{(1)}_{\mu\nu}&=&A_{1}{\partial\over\partial s}+B_{1}{\partial\over\partial t}, 
\nonumber \\[0.2cm]
\epsilon^{\mu}k^{\nu}J^{(2')}_{\mu\nu}&=&A_{2'}{\partial\over\partial s}+B_{2'}{\partial\over\partial t}, 
\\[0.2cm]
\epsilon^{\mu}k^{\nu}J^{(2)}_{\mu\nu}&=&A_{2}{\partial\over\partial s}+B_{2}{\partial\over\partial t},
\nonumber
\end{eqnarray}
where the gauge invariant coefficients $A_{i}$ and $B_{i}$ are defined by
\begin{eqnarray}
A_{1'}&=&(\epsilon\cdot p')(q'\cdot k)-(\epsilon\cdot q')(p'\cdot k), \nonumber \\[0.2cm]
A_{1}&=&(\epsilon\cdot p)(q\cdot k)-(\epsilon\cdot q)(p\cdot k),
\label{eq:A's_coeff_def} \\[0.2cm]
A_{2'}&=&(\epsilon\cdot q')(p'\cdot k)-(\epsilon\cdot p')(q'\cdot k), \nonumber \\[0.2cm]
A_{2}&=&(\epsilon\cdot q)(p\cdot k)-(\epsilon\cdot p)(q\cdot k), \nonumber
\end{eqnarray}
and
\begin{eqnarray}
B_{1'}&=&(\epsilon\cdot p)(p'\cdot k)-(\epsilon\cdot p')(p\cdot k), \nonumber \\[0.2cm]
B_{1}&=&(\epsilon\cdot p')(p\cdot k)-(\epsilon\cdot p)(p'\cdot k), \label{eq:B's_coeff_def} \\[0.2cm]
B_{2'}&=&(\epsilon\cdot q)(q'\cdot k)-(\epsilon\cdot q')(q\cdot k), \nonumber \\[0.2cm]
B_{2}&=&(\epsilon\cdot q')(q\cdot k)-(\epsilon\cdot q)(q'\cdot k). \nonumber 
\end{eqnarray}
Plugging these expressions in the amplitude, we arrive at
\begin{eqnarray}
\mathcal{A}_{5}&=&2g \Bigg[c_{1}{p'\cdot\epsilon\over s_{1'}}-c_{2}{p\cdot\epsilon\over s_{1}}
+c_{4}{q'\cdot \epsilon\over s_{2'}}-c_{5}{q\cdot \epsilon\over s_{2}} 
\nonumber \\[0.2cm]
&+&\left(c_{1}{A_{1'}\over s_{1'}}+c_{2}{A_{1}\over s_{1}}+c_{4}{A_{2'}\over s_{2'}}+c_{5}{A_{2}\over s_{2}}\right)
{\partial\over\partial s}
\label{eq:gauge_amplitude_low_st_first} \\[0.2cm]
&+&\left(c_{1}{B_{1'}\over s_{1'}}+c_{2}{B_{1}\over s_{1}}+c_{4}{B_{2'}\over s_{2'}}+c_{5}{B_{2}\over s_{2}}\right)
{\partial\over\partial t} \Bigg] \mathcal{A}_{4}(s,t).
\nonumber
\end{eqnarray}

In this expression gauge invariance follows trivially from the invariance of the coefficients $A_{i}$ and $B_{i}$. It is important to note that there exists a larger set of transformations of these coefficients that leaves the first correction to Low's theorem invariant. These are given by the shifts
\begin{eqnarray}
A_{1'}&\longrightarrow& A_{1'}+s_{1'} \, \alpha(p,q,p',q'), \nonumber \\[0.2cm]
A_{1}&\longrightarrow& A_{1}-s_{1} \, \alpha(p,q,p',q'), \nonumber \\[0.2cm]
A_{2'}&\longrightarrow& A_{2'}+s_{2'} \, \alpha(p,q,p',q'), \\[0.2cm]
A_{2}&\longrightarrow& A_{2}-s_{2} \, \alpha(p,q,p',q'), \nonumber
\end{eqnarray}
and
\begin{eqnarray}
B_{1'}&\longrightarrow& B_{1'}+s_{1'} \, \beta(p,q,p',q'), \nonumber \\[0.2cm]
B_{1}&\longrightarrow& B_{1}-s_{1} \, \beta(p,q,p',q'), \nonumber \\[0.2cm]
B_{2'}&\longrightarrow& B_{2'}+s_{2'} \, \beta(p,q,p',q'), \\[0.2cm]
B_{2}&\longrightarrow& B_{2}-s_{2} \, \beta(p,q,p',q'), \nonumber
\end{eqnarray}
where $\alpha(p,q,p',q')$ and $\beta(p,q,p',q')$ are two arbitrary functions of the scalar momenta, not necessarily local. Note that these transformations resemble those of the original color-kinematics duality~\cite{BCJ}, although they are only affecting the factorizing soft factors and not the full amplitude. Moreover, we have not identified any relevant role for the Jacobi identities when investigating  the double copy structure in the gravitational case. This is likely to be a feature of the soft limit alone.

\section{The gravitational amplitude and its double copy structure}
\label{sec:grav_amplitu_dc}

The amplitude for the scattering of two distinct scalars with emission of a graviton can be computed in the 
soft limit by considering the five generic topologies shown in Fig.~\ref{fig:topologies} with the gluon replaced by a graviton. Following the general arguments given in~\cite{BDDVN}, the result is 
\begin{eqnarray}
\mathcal{M}_{5}&=&{\kappa} \left(-{p\cdot\varepsilon\cdot p\over s_{1}}+{p'\cdot\varepsilon\cdot p'\over s_{1'}}
-{q\cdot\varepsilon\cdot q\over s_{2}}+{q'\cdot \varepsilon\cdot q'\over s_{2'}}\right.
\label{eq:grav_amplitude}\\[0.2cm]
&+&\left.{p'_{\mu}\varepsilon^{\mu\nu}k^{\alpha}J^{(1')}_{\nu\alpha}\over s_{1'}}
+{p_{\mu}\varepsilon^{\mu\nu}k^{\alpha}J^{(1)}_{\nu\alpha}\over s_{1}}
+{q'_{\mu}\varepsilon^{\mu\nu}k^{\alpha}J^{(2')}_{\nu\alpha}\over s_{2'}}
+{q_{\mu}\varepsilon^{\mu\nu}k^{\alpha}J^{(2)}_{\nu\alpha}\over s_{2}}
\right)\mathcal{M}_{4}(p,q,p',q'),
\nonumber
\end{eqnarray}
where $\mathcal{M}_{4}(p,q,p',q')$ denotes the gravitational scattering amplitude of the four hard (scalar) particles. The subleading term can 
be recast in terms of derivatives with respect to the kinematic invariants $s$ and $t$ using
\begin{eqnarray}
p'_{\mu}\varepsilon^{\mu\nu}k^{\alpha}J^{(1')}_{\nu\alpha}
&=&\widetilde{A}_{1'}{\partial\over\partial s}+\widetilde{B}_{1'}{\partial\over\partial t},
\nonumber \\[0.2cm]
p_{\mu}\varepsilon^{\mu\nu}k^{\alpha}J^{(1)}_{\nu\alpha}
&=&\widetilde{A}_{1}{\partial\over\partial s}+\widetilde{B}_{1}{\partial\over\partial t},
\nonumber \\[0.2cm]
q'_{\mu}\varepsilon^{\mu\nu}k^{\alpha}J^{(2')}_{\nu\alpha}&=&
\widetilde{A}_{2'}{\partial\over\partial s}+\widetilde{B}_{2'}{\partial\over\partial t},
\\[0.2cm]
{q_{\mu}\varepsilon^{\mu\nu}k^{\alpha}}J^{(2)}_{\nu\alpha}&=&
\widetilde{A}_{2}{\partial\over\partial s}+\widetilde{B}_{2}{\partial\over\partial t},
\nonumber
\end{eqnarray}
where the new coefficients $\widetilde{A}_{i}$ and $\widetilde{B}_{i}$ are given by
\begin{eqnarray}
\widetilde{A}_{1'}&=&(p'\cdot\varepsilon\cdot p')(q'\cdot k)-
(p'\cdot \varepsilon\cdot q')(p'\cdot k),\nonumber \\[0.2cm]
\widetilde{A}_{1}&=&(p\cdot\varepsilon\cdot p)(q\cdot k)-(p\cdot\varepsilon\cdot q)(p\cdot k), \nonumber \\[0.2cm]
\widetilde{A}_{2'}&=&(q'\cdot\varepsilon\cdot q')(p'\cdot k)-(q'\cdot\varepsilon\cdot p')(q'\cdot k), \\[0.2cm]
\widetilde{A}_{2} &=&(q\cdot\varepsilon\cdot q)(p\cdot k)-(q\cdot\varepsilon\cdot p)(q\cdot k),
\nonumber
\end{eqnarray}
and
\begin{eqnarray}
\widetilde{B}_{1'}&=&(p'\cdot\varepsilon\cdot p)(p'\cdot k)-(p'\cdot\varepsilon\cdot p')(p\cdot k), \nonumber \\[0.2cm]
\widetilde{B}_{1}&=&(p\cdot\varepsilon\cdot p')(p\cdot k)-(p\cdot\varepsilon\cdot p)(p'\cdot k), \nonumber \\[0.2cm]
\widetilde{B}_{2'}&=&(q'\cdot\varepsilon\cdot q)(q'\cdot k)-(q'\cdot\varepsilon\cdot q')(q\cdot k), \\[0.2cm]
\widetilde{B}_{2}&=&(q\cdot\varepsilon\cdot q')(q\cdot k)-(q\cdot\varepsilon\cdot q)(q'\cdot k).
\nonumber
\end{eqnarray}
In terms of this set of gauge invariant coefficients,  
the amplitude \eqref{eq:grav_amplitude} reads
\begin{eqnarray}
\mathcal{M}_{5}&=&{\kappa}\Bigg[-{p\cdot\varepsilon\cdot p\over s_{1}}+{p'\cdot\varepsilon\cdot p'\over s_{1'}}
-{q\cdot\varepsilon\cdot q\over s_{2}}+{q'\cdot \varepsilon\cdot q'\over s_{2'}}
\label{eq:grav_amplitude_low_st_first} \\[0.2cm]
&+&\left({\widetilde{A}_{1'}\over s_{1'}}+{\widetilde{A}_{1}\over s_{1}}+{\widetilde{A}_{2'}\over s_{2'}}
+{\widetilde{A}_{2}\over s_{2}}\right){\partial\over\partial s}
+\left({\widetilde{B}_{1'}\over s_{1'}}+{\widetilde{B}_{1}\over s_{1}}+{\widetilde{B}_{2'}\over s_{2'}}
+{\widetilde{B}_{2}\over s_{2}}\right){\partial\over\partial t}
\Bigg] \mathcal{M}_{4}(s,t). \nonumber
\end{eqnarray}

Similarly to the gauge theory case, the gravitational amplitude also remains invariant under the following generalized transformations of the coefficients (with $i=1,1',2,2'$): 
\begin{eqnarray}
\widetilde{A}_{i}&\longrightarrow& \widetilde{A}_{i}+s_{i} \, \widetilde{\alpha}_{i}(p,q,p',q'),  \nonumber \\[0.2cm]
\widetilde{B}_{i}&\longrightarrow& \widetilde{B}_{i}+s_{i} \, \widetilde{\beta}_{i}(p,q,p',q'),
\label{eq:generalized_gauge_t_grav}
\end{eqnarray}
where the functions $\widetilde{\alpha}_{i}(p,q,p',q')$ and $\widetilde{\beta}_{i}(p,q,p',q')$ satisfy the constraint
\begin{eqnarray}
\sum_{i}\widetilde{\alpha}_{i}(p,q,p',q')=0, \hspace*{1cm} \sum_{i}\widetilde{\beta}_{i}(p,q,p',q')=0. 
\end{eqnarray}
These transformations turn out to be useful in finding a relation between the gravitational and gauge theory amplitudes. Transforming 
$\widetilde{A}_{i}$ and $\widetilde{B}_{i}$ using the functions
\begin{eqnarray}
\widetilde{\alpha}_{1'}&=&- \widetilde{\alpha}_{2'} ~=~ {-(p'\cdot\varepsilon\cdot p')(q'\cdot k)+(p'\cdot\varepsilon\cdot q')[(p'-q')\cdot k]+(q'\cdot\varepsilon\cdot q')(p'\cdot k)
\over 2k\cdot (p'+q')}, \nonumber \\[0.2cm]
\widetilde{\alpha}_{1}&=&- \widetilde{\alpha}_{2} ~=~{-(p\cdot\varepsilon\cdot p)(q\cdot k)+(p\cdot\varepsilon\cdot q)[(p-q)\cdot k]
+(q\cdot\varepsilon\cdot q)(p\cdot k)\over 2k\cdot (p+q)}, \nonumber \\[0.2cm]
\widetilde{\beta}_{1'}&=&- \widetilde{\beta}_{1} ~=~{-(p'\cdot\varepsilon\cdot p')(p\cdot k)+(p'\cdot\varepsilon\cdot  p)[k\cdot (p+p')]
-(p\cdot\varepsilon\cdot p)(p'\cdot k)\over 2k\cdot (p-p')}, \\[0.2cm]
\widetilde{\beta}_{2'}&=&- \widetilde{\beta}_{2} ~=~{-(q'\cdot\varepsilon\cdot q')(q\cdot k)+(q\cdot\varepsilon\cdot q')[k\cdot (q+q')]
-(q\cdot\varepsilon\cdot q)(q'\cdot k)\over 2k\cdot (q-q')}, \nonumber
\end{eqnarray}
we find the new coefficients ${\widetilde A}'_{i}$ and ${\widetilde B}'_{i}$ given by
\begin{eqnarray}
\widetilde{A}_{1'}&\longrightarrow & \widetilde{A}_{1'}' ~=~2
{(p'\cdot \varepsilon\cdot p')(q'\cdot k)^{2}-2(p'\cdot\varepsilon\cdot q')(p'\cdot k)(q'\cdot k)+(q'\cdot\varepsilon\cdot q')
(p'\cdot k)^{2}\over s_{1'}+s_{2'}}, \nonumber \\[0.2cm]
\widetilde{A}_{1}&\longrightarrow & \widetilde{A}_{1}' ~=~ 2{(p\cdot\varepsilon\cdot p)(p\cdot k)^{2}
-2(p\cdot\varepsilon\cdot q)(p\cdot k)(q\cdot k)
+(q\cdot \varepsilon\cdot q)(q\cdot k)^{2}\over s_{1}+s_{2}}, \\[0.2cm]
\widetilde{A}_{2'}&\longrightarrow & \widetilde{A}'_{2'} ~=~ 2{(p'\cdot\varepsilon\cdot p')(q'\cdot k)^{2}
-2(p'\cdot\varepsilon\cdot q')(p'\cdot k)(q'\cdot k)+(q'\cdot\varepsilon\cdot q')(p'\cdot k)^{2}\over s_{1'}+s_{2'}},
\nonumber \\[0.2cm]
\widetilde{A}_{2}&\longrightarrow& \widetilde{A}'_{2} ~=~ 2{(p\cdot\varepsilon\cdot p)(q\cdot k)^{2}
-2(p\cdot\varepsilon\cdot q)(p\cdot k)(q\cdot k)+(q\cdot\varepsilon\cdot q)(p\cdot k)^{2}\over s_{1}+s_{2}},
\nonumber 
\end{eqnarray}
and
\begin{eqnarray}
\widetilde{B}_{1'}&\longrightarrow & \widetilde{B}_{1'}' ~=~ 
-2{(p'\cdot\varepsilon\cdot p')(p'\cdot k)^{2}-2(p'\cdot\varepsilon\cdot p)(p\cdot k)(p'\cdot k)
+(p\cdot\varepsilon\cdot p)(p'\cdot k)^{2}\over t_{1}-t_{2}},
\nonumber \\[0.2cm]
\widetilde{B}_{1}&\longrightarrow & \widetilde{B}_{1}' ~=~ 2{(p\cdot\varepsilon\cdot p)(p\cdot k)^{2}
-2(p\cdot\varepsilon\cdot p')(p\cdot k)(p'\cdot k)+(p'\cdot\varepsilon\cdot p')(p\cdot k)^{2}\over t_{1}-t_{2}},
 \\[0.2cm]
\widetilde{B}_{2'}&\longrightarrow & \widetilde{B}'_{2'} ~=~ 2{(q'\cdot\varepsilon\cdot q')(q\cdot k)^{2}
-2(q\cdot\varepsilon\cdot q')(q\cdot k)(q'\cdot k)+(q\cdot\varepsilon\cdot q)(q'\cdot k)^{2}
\over t_{1}-t_{2}},
\nonumber \\[0.2cm]
\widetilde{B}_{2}&\longrightarrow & \widetilde{B}'_{2} ~=~ -2{(q\cdot\varepsilon\cdot q)(q'\cdot k)^{2}
-2(q\cdot\varepsilon\cdot q')(q\cdot k)(q'\cdot k)+(q'\cdot \varepsilon\cdot q')(q\cdot k)^{2}\over
t_{1}-t_{2}}.
\nonumber
\end{eqnarray}
Here, to simplify the notation, we have introduced the invariants
\begin{eqnarray}
t_{1}=(p-p')^{2}, \hspace*{1cm} t_{2}=(q-q')^{2}.
\end{eqnarray}
This transformation is interesting because if we compare the new set of coefficients with the $A_{i}$'s and $B_{i}$'s of the gauge theory amplitude given 
in Eqs. \eqref{eq:A's_coeff_def} and \eqref{eq:B's_coeff_def} we find the relations
\begin{eqnarray}
\widetilde{A}'_{1'}&=&{2\varepsilon_{\mu\nu}A_{1'}^{\mu}A_{1'}^{\nu}\over s_{1'}+s_{2'}}, \nonumber\\[0.2cm]
\widetilde{A}'_{1}&=&{2\varepsilon_{\mu\nu}A_{1}^{\mu}A_{1}^{\nu}\over s_{1}+s_{2}}, \\[0.2cm]
\widetilde{A}'_{2'}&=&{2\varepsilon_{\mu\nu}A_{2'}^{\mu}A_{2'}^{\nu}\over s_{1'}+s_{2'}}, \nonumber \\[0.2cm]
\widetilde{A}'_{2}&=&{2\varepsilon_{\mu\nu}A_{2}^{\mu}A_{2}^{\nu}\over s_{1}+s_{2}}
\nonumber
\end{eqnarray}
and
\begin{eqnarray}
\widetilde{B}'_{1'}&=&-{2\varepsilon_{\mu\nu}B_{1'}^{\mu}B_{1'}^{\nu}\over t_{1}-t_{2}}, \nonumber\\[0.2cm]
\widetilde{B}'_{1}&=&{2\varepsilon_{\mu\nu}B_{1}^{\mu}B_{1}^{\nu}\over t_{1}-t_{2}}, \\[0.2cm]
\widetilde{B}'_{2'}&=&{2\varepsilon_{\mu\nu}B_{2'}^{\mu}B_{2'}^{\nu}\over t_{1}-t_{2}}, \nonumber \\[0.2cm]
\widetilde{B}'_{2}&=&-{2\varepsilon_{\mu\nu}B_{2}^{\mu}B_{2}^{\nu}\over t_{1}-t_{2}}.
\nonumber
\end{eqnarray}
Thus, up to a common kinematic denominator and a phase (which can be absorbed in a redefinition of the $B_{i}$'s), 
the coefficients of the gravity amplitude can be written 
as a double copy of the ones of the gauge theory. This structure is manifest if we rewrite the
scalar amplitude in the gauge theory
\begin{eqnarray}
\mathcal{A}_{5}&=&2\, g \, \epsilon_{\mu}\left[c_{1}{p'^\mu \over s_{1'}}-c_{2}{p^\mu \over s_{1}}
+c_{4}{q'^\mu \over s_{2'}}-c_{5}{q^\mu \over s_{2}}
\right.
\label{eq:gauge_amplitude_epsilon_factorized}\\[0.2cm]
&+& \left(c_{1}{A^{\mu}_{1'}\over s_{1'}+s_{2'}}{1\over s_{1'}}+c_{2}{A_{1}^{\mu}\over s_{1}+s_{2}}{1\over s_{1}}
+c_{4}{A_{2'}^{\mu}\over s_{1'}+s_{2'}}{1\over s_{2'}}+c_{5}{A_{2}^{\mu}\over s_{1}+s_{2}}{1\over s_{2}}\right)
(s_{1}+s_{2}){\partial\over\partial s}
\nonumber\\[0.2cm]
&+& \left.\left(c_{1}{B_{1'}^{\mu}\over t_{1}-t_{2}}{1\over s_{1'}}+c_{2}{B_{1}^{\mu}\over t_{1}-t_{2}}{1\over s_{1}}
+c_{4}{B^{\mu}_{2'}\over t_{1}-t_{2}}{1\over s_{2'}}+c_{5}{B^{\mu}_{2}\over t_{1}-t_{2}}{1\over s_{2}}\right)
(t_{1}-t_{2}){\partial\over\partial t} \right] \mathcal{A}_{4}(s,t) \nonumber 
\end{eqnarray}
and compare with the gravitational amplitude written as
\begin{eqnarray}
\mathcal{M}_{5}&=&{\kappa} \, \varepsilon_{\mu\nu}\left\{{p'^{\mu} p'^{\nu}\over s_{1'}}
- {p^{\mu} p^{\nu} \over s_{1}}
+{q'^{\mu} q'^{\nu}  \over s_{2'}}-{q^{\mu} q^{\nu} \over s_{2}}
\right.
\label{eq:gravitational_amplitude_epsilon_factorized}\\[0.2cm]
&+&2 \left[{A_{1'}^{\mu}A_{1'}^{\nu}\over (s_{1'}+s_{2'})^{2}}{1\over s_{1'}}+
{A_{1}^{\mu}A_{1}^{\nu}\over (s_{1}+s_{2})^{2}}{1\over s_{1}}
+{A_{2'}^{\mu}A_{2'}^{\nu}\over (s_{1'}+s_{2'})^{2}}{1\over s_{2'}}
+{A_{1}^{\mu}A_{1}^{\nu}\over (s_{1}+s_{2})^{2}}{1\over s_{2}}\right](s_{1}+s_{2}){\partial\over\partial s}
\nonumber \\[0.2cm]
&+&\left. 2\left[-{B_{1'}^{\mu}B_{1'}^{\nu}\over (t_{1}-t_{2})^{2}}{1\over s_{1'}}
+{B_{1}^{\mu}B_{1}^{\nu}\over (t_{1}-t_{2})^{2}}{1 \over s_{1}}
+{B_{2'}^{\mu}B_{2'}^{\nu} \over (t_{1}-t_{2})^{2}}{1\over s_{2'}}
-{B_{1}^{\mu}B_{1}^{\nu}\over (t_{1}-t_{2})^{2}}{1\over s_{2}}
\right](t_{1}-t_{2}){\partial\over\partial t}\right\}\mathcal{M}_{4}(s,t).
\nonumber
\end{eqnarray}
We have used momentum conservation $s_{1}+s_{2}=s_{1'}+s_{2'}$. 

Some remarks on the expressions \eqref{eq:gauge_amplitude_epsilon_factorized} and \eqref{eq:gravitational_amplitude_epsilon_factorized} are in order. 
Factoring out $s_{1}+s_{2}$ and $t_{1}-t_{2}$ might seem a mere analytic trick to get the double copy to work better. However, this way of writing 
the amplitudes is quite natural once we take into account that these two terms are the expansion parameters in the soft limit around $s$ and $t$, so the
double copy representation affects the coefficients of the series expansion around the $k=0$ term.
Written in this way, it is clear how the contribution of the soft graviton to the five-point amplitude can be obtained by replacing the color factors in the
gauge amplitude with a second copy of the corresponding kinetic coefficient. This prescription works not only for the correction but for the leading term as 
well, where the kinematic coefficient is just the momentum.

We should also stress the importance of using derivatives with respect to the kinematic invariants in uncovering the double copy structure of the subleading corrections in the soft limit. In this case, the tensor structure of the amplitude is completely 
codified in the coefficients of these derivatives, in which the double copy is glaring. Expressing the amplitude in terms of 
derivatives with respect to the momenta obscures this feature. 

\section{Gribov's limit}
\label{sec:gribov}

We have considered so far amplitudes in the standard soft gluon and graviton regimes, as well as their first corrections. 
As explained in the Introduction, Gribov found
that Low's result is valid in a kinematic region 
larger than the strict $k^{\mu}\rightarrow 0$ limit. In our notation this is given by
\begin{eqnarray}
s_{1},s_{2}\ll s, \hspace*{1cm} \mathbf{k}_{\perp}^{2}\ll \mu^{2}\ll s,
\label{eq:gribov2_s1_s2}
\end{eqnarray}
with $\mu$ the mass of the scalars. Moreover, in this limit it is also satisfied \cite{gribov}
\begin{eqnarray}
|t_{1}-t_{2}|\ll \mu\sqrt{-t_{1}}\approx \mu\sqrt{-t_{2}}.
\label{eq:gribov_t1-t2}
\end{eqnarray}

It is remarkable that in this region the associated radiation can be hard, {\it i.e.}, we are not just limited to emission of soft particles. Gribov's result was extended in~\cite{lipatov1} to the scattering amplitude of two scalar flavors in nonabelian gauge theories 
in the high energy limit $s\gg t\sim \mu^{2}$. This inequality has important consequences for the form of the amplitude. Since
the four-scalar amplitude $\mathcal{A}_{4}(s,t)$ is dimensionless, it has to be a homogeneous function of degree zero of its two arguments. 
This means that in this kinematic regime, derivatives with respect to $s$ are much smaller than the derivatives with respect to $t$, due to a suppression factor $t/s$: 
\begin{eqnarray}
\left(s{\partial\over\partial s}+t{\partial\over\partial t}\right)\mathcal{A}_{4}(s,t)=0 \hspace*{1cm} \Longrightarrow \hspace*{1cm}
{\partial\over\partial s}\mathcal{A}_{4}(s,t)=-{t\over s}{\partial\over\partial t}\mathcal{A}_{4}(s,t).
\end{eqnarray}
As a consequence, writing Eq.~\eqref{eq:general_amplitude_proof} in terms of the kinematic invariants, the momentum shifts in the expression only affect the second argument of $\mathcal{A}(s,t)$, {\it i.e.},
\begin{eqnarray}
\mathcal{A}_{5}&=&2g\left[c_{1}{p'\cdot\epsilon\over s_{1'}}\mathcal{A}_{4}(s,t_{2})-c_{2}{p\cdot\epsilon\over s_{1}}
\mathcal{A}_{4}(s,t_{2})+c_{4}{q'\cdot \epsilon\over s_{2'}}\mathcal{A}_{4}(s,t_{1})-c_{5}{q\cdot \epsilon\over s_{2}}
\mathcal{A}_{4}(s,t_{1})\right]  \\[0.2cm]
&+& {g\over 2} \epsilon_{\mu} \Big[c_{3} \, \mathcal{B}_{1}^{\mu}(k;p,q,p',q')
+ c_{6} \, \mathcal{B}_{2}^{\mu}(k;p,q,p',q')
+ c_{7} \, \mathcal{B}_{3}^{\mu}(k;p,q,p',q')\Big].
\nonumber
\end{eqnarray}
Again, to determine the unknown functions associated with internal gluon emission we write the gauge Ward identity and take into
account that in our kinematic regime 
\begin{eqnarray}
{t_{1}-t_{2}\over 2}\ll {t_{1}+t_{2}\over 2}\equiv t.
\end{eqnarray}
Then, we expand the expression to first order in $t_{1}-t_{2}$. 
Since this parameter is proportional to $k$, the leading contribution of the 
functions $\mathcal{B}_{i}^{\mu}(k;p,q,p',q')$ to the Ward identity comes from setting $k=0$ in the argument. Proceeding as in
Section~\ref{sec:soft_QCD}, we find that gauge invariance fixes the unknown functions and the result of~\cite{lipatov1} is recovered
in a slightly different notation:
\begin{eqnarray}
\mathcal{A}_{5}&=&2 g \Bigg[c_{1}{p'\cdot\epsilon\over s_{1'}}-c_{2}{p\cdot\epsilon\over s_{1}}
+c_{4}{q'\cdot \epsilon\over s_{2'}}-c_{5}{q\cdot \epsilon\over s_{2}}
\nonumber \\[0.2cm]
&+&\left(c_{1}{B_{1'}\over s_{1'}}+c_{2}{B_{1}\over s_{1}}+c_{4}{B_{2'}\over s_{2'}}+c_{5}{B_{2}\over s_{2}}\right)
{\partial\over\partial t} \Bigg]\mathcal{A}_{4}(s,t).
\label{eq:gauge_gribov_limit}
\end{eqnarray}

A similar calculation can be carried out for the gravitational amplitude. The only caveat lies in whether we can neglect derivatives
with respect to $s$, since now due to the dimensionful coupling $\kappa$ it is no longer true that the amplitude is a function of $s/t$. 
Nevertheless, at a fixed order in perturbation theory the amplitude has the generic form
\begin{eqnarray}
\mathcal{M}_{4}(s,t)=(\kappa^{2} s)^{n\over 2}f\left({s\over t}\right).
\end{eqnarray}
However, if at large energies $f(s/t)\sim (s/t)^{\alpha}$, the $s$-derivative of $\mathcal{M}_{4}(s,t)$ is suppressed with respect to its 
$t$-derivatives by a power of $t/s$. This is indeed the case of the tree-level amplitude (with $\alpha=1$), so
we can take the amplitude as constant with respect to $s$ and retrieve the expression found
in~\cite{lipatov1}:
\begin{eqnarray}
\mathcal{M}_{5}&=&{\kappa} \Bigg[{p'\cdot\varepsilon\cdot p'\over s_{1'}}-{p\cdot\varepsilon\cdot p\over s_{1}}
+{q'\cdot \varepsilon\cdot q'\over s_{2'}} -{q\cdot\varepsilon\cdot q\over s_{2}}
\nonumber \\[0.2cm]
&+& \left({\widetilde{B}_{1'}\over s_{1'}}+{\widetilde{B}_{1}\over s_{1}}+{\widetilde{B}_{2'}\over s_{2'}}
+{\widetilde{B}_{2}\over s_{2}}\right){\partial \over\partial t} \Bigg] \mathcal{M}_{4}(s,t). 
\label{eq:grav_gribov_limit}
\end{eqnarray}
We see how Gribov's limit of the gauge and gravitational scattering amplitudes \eqref{eq:gauge_gribov_limit} and \eqref{eq:grav_gribov_limit}
 can be respectively obtained from our expressions for the corrections to the
Low and Weinberg limits \eqref{eq:gauge_amplitude_low_st_first} and \eqref{eq:grav_amplitude_low_st_first} by just ignoring derivatives with respect to $s$.

\section{Concluding remarks}
\label{sec:conclusions}

The idea of the existence of a double copy representation of gravity has received strong support, ranging from KLT identities~\cite{KLT} to color-kinematics 
duality~\cite{BCJ} (see~\cite{elvang_huang}
for a recent review). In the context of the soft limit, it was found in~\cite{oxburgh_white} that the infrared behavior of both gauge theories and
gravity is consistent with an underlying double copy provided by color-kinematics duality to all orders in perturbation theory.  

In this paper we have studied the double copy structure in the context of the soft gluon and graviton theorems. Our analysis shows clear evidence that there is a sense in which we can state that (soft graviton) = (soft gluon)$^{2}$: the contribution of a soft graviton in a scalar scattering amplitude can be written as the double copy of the corresponding contribution of a soft gluon. 

Let us try to be more precise. Our proposal strongly resembles color-kinematics duality of gauge theory amplitudes, in which the gravity amplitude is obtained by replacing color factors by a second copy of a kinematic factor. It has however the peculiarity that it does not affect the whole five-point amplitude, but just the coefficients of the operator acting on 
the amplitude of the four hard particles. The rationale behind this is that it is this prefactor which contains all the information about the emitted
gluon/graviton. This is precisely the sense of the moral equation (soft graviton)=(soft gluon)$^{2}$. 

Interestingly, in the case of the scalar QCD five-point amplitude studied here, it was shown in~\cite{SVSCVM2} that a naive application of color-kinematics
duality does not render the full gravitational amplitude for graviton emission. Even in multi-Regge kinematics, 
the full graviton amplitude does not factorize in terms of two QCD effective gluon vertices (Lipatov vertices), 
due to an extra term which is necessary for the cancellation 
of overlapping divergences \cite{SVSCVM}. Despite this, here we have seen how a certain double copy
structure does indeed survive in the Low/Weinberg and Gribov limits, but that this only affects the soft graviton. Incidentally, the 
soft graviton limit does not contain overlapping divergences, so the offending term breaking factorization in the Regge limit does not contribute to it. Let us remark that by going to the Gribov limit it is possible to escape from the soft graviton condition and extend Low's factorization to the emission of harder radiation. This is likely to have non-trivial consequences for the interpretation of the Gribov limit in terms of symmetries in a gravitational theory at null infinity.

The scalar QCD theory used here admits modifications for which the Bern-Carrasco-Johansson prescription in a general kinematics works~\cite{JSVSCVM}. One of them consists in taking all scalars to transform 
in the adjoint representation and adding a quartic contact self-coupling between the two flavors. Nevertheless, this modification does not spoil
the double copy structure of the subleading terms, since the new diagrams only add to the leading soft behavior. This is because the new couplings
are contact terms, so they only contribute to the sector with zero angular momentum for which the correction vanishes.

There are a number of related questions that deserve attention. One of them concerns the generalization of our result to other amplitudes and theories at tree and loop level ({\it e.g.}, 
is there a hidden double copy interpretation of  the double logarithms studied in~\cite{Bartels:2012ra}?). 
In this sense, as already stated in Section \ref{sec:grav_amplitu_dc}, the use of derivatives with respect to invariants seems crucial to expose the double
copy structure in the amplitude. This is an added complication when investigating higher-point amplitudes. Another
interesting problem to address is the implications of our results for the possible relation between the asymptotic symmetries of gauge theories
\cite{lysov_et_al,lysov_et_al2} and gravity \cite{strominger_et_al}. These and other issues will be studied elsewhere.

\section*{Acknowledgments}

A.S.V. acknowledges support from European Commission under contract LHCPhenoNet (PITN-GA-2010-264564), Madrid Regional Government (HEPHACOSESP-1473), Spanish Government (MICINN (FPA2010-17747)) and Spanish MINECO Centro de Excelencia Severo Ochoa Programme (SEV-2012-0249). The work of M.A.V.-M. has been partially supported by Spanish Government grants FPA2012-34456 and FIS2012-30926, Basque Government Grant IT-357-07 and Spanish Consolider-Ingenio 2010 Programme CPAN (CSD2007-00042). He also thanks the Instituto de F\'{\i}sica Te\'orica UAM/CSIC for hospitality during the completion of this work.

\end{document}